\documentclass[aps,preprint]{revtex4}
\usepackage{graphicx}
\usepackage{dcolumn}
\usepackage{bm}
\usepackage{hyperref}
\usepackage{subfigure}
\usepackage[sort&compress]{natbib}
\begin{document}

\title{Crinkles in the last scattering surface: Non-Gaussianity from
  inhomogeneous recombination.}
\author{Rishi Khatri}
\email{rkhatri2@illinois.edu}
\affiliation{Department of Astronomy, University of Illinois at Urbana-Champaign, 1002 W.~Green Street, Urbana, IL 61801}
\author{Benjamin D. Wandelt}
\email{bwandelt@illinois.edu}
\affiliation{Departments of Physics and Astronomy, University of Illinois at Urbana-Champaign, 1002 W.~Green Street, Urbana, IL 61801}

\date{\today}
\begin{abstract}
  The perturbations in the electron number density during recombination
  contributes to the cosmic microwave background bispectrum through second
  order terms. Perturbations in the electron density can be a factor of
  $\sim 5$ larger than the baryon density fluctuations on large scales as
  shown in the calculations by  Novosyadlyj. This raises the possibility that
  the contribution to the bispectrum arising from perturbations in the optical
  depth may be non-negligible. We calculate this bispectrum and find it to
  peak for squeezed triangles and of peak amplitude of the order of 
  primordial non-Gaussianity of local type with $f_{NL}\approx 0.05 \sim -1$
  depending on the $\ell$-modes being considered. This is because the shape of
  the bispectrum is different  from the primordial one although it peaks
  for squeezed configurations, similar to the local type primordial
  non-Gaussianity.
 \end{abstract}
\maketitle
\begin{section}{Introduction}
First order perturbation theory has been of sufficient accuracy  for
analysis of the Cosmic Microwave
Background (CMB) observations so far. However future CMB experiments will
have high enough precision that second order effects would need to be taken
into account for theory to have similar accuracy. The second order
contributions will in particular be important for the higher order statistics
like the three point correlation or the bispectrum. Second order effects
in CMB  have been studied previously \cite{0,1,2,3,4,5,6,7,8,9,10,11,12}.
Bartolo et al. have derived the Boltzmann equations  at second order and
also the analytic solutions for the CMB transfer function at second order
with some simplifying assumptions \cite{bar1,bar2}, see also \cite{pitrou}.

All numerical and analytic calculations at second order so far have
ignored the contribution arising from the perturbations in the electron
number density, $\delta_e$. These contributions are expected to be small
compared to other second order terms since $\delta_e$ multiplies the
collision term which has contributions from the difference of first order
radiation and electron dipoles , radiation quadrupole and  higher
order moments of the radiation transfer function. These terms are small
during recombination compared to the monopole terms.
Recombination however depends on matter and radiation densities  and
perturbations in the
electron number density can be quite different from the perturbations in
the matter and radiation densities. This was calculated by Novosyadlyj
\cite{novos} who showed that this is indeed the case and perturbations in
the electron number density can be $\sim 5$ times the baryon number density
perturbations. 

We calculate the CMB bispectrum on all scales
arising due to the perturbations
in electron number density and compare
it with the bispectrum expected from a primordial non-Gaussianity of the
local type. This is a full numerical calculation without any other
approximation except  that we only consider terms involving
$\delta_e$. Although this bispectrum turns out to be below the 
detection levels of future experiments like Planck \cite{planck}, there are
some important general implications which are discussed in the conclusions
section. We use the following cosmological parameters for our calculations 
(values at redshift $z=0$ unless specified):
baryon density $\Omega_b=0.0418$, cold dark matter density
$\Omega_c=0.19647$, cosmological constant $\Omega_{\Lambda}=0.76173$,
number of massless neutrinos $N_{\nu}=3.04$, Hubble constant $H_0=73$, CMB
temperature $T_{CMB}=2.725$, primordial Helium fraction $y_{He}=0.24$,
redshift of reionization $z_{ri}=10$, primordial gravitational potential power spectrum $P(k)=2\pi^2/k^3$

\end{section}

\begin{section}{Inhomogeneous Recombination}
We use the code DRECFAST \cite{drecfast} by
Novosyadlyj \cite{novos}, which is a modification of the recombination code
RECFAST \cite{recfast} to calculate the perturbations in the electron
number density $\delta_e=(n_e-\bar{n_e})/\bar{n_e}$ during
recombination. $n_e$ is the local electron number density and $\bar{n_e}$
is the mean electron density.  Perturbations in baryon ($\delta_b$) and photon density
($\delta_{\gamma}$) result in perturbations
in the electron density with an amplitude that is amplified or suppressed
depending on which terms in the evolution equations prevail. Specifically
photo-ionization prevails on superhorizon scales resulting in
$\delta_e \sim 5\times \delta_b$  during recombination.  We  refer the
reader to \cite{novos} for further details. 

We will not consider the full second order Boltzmann equation \cite{bar1}
but only the terms involving the perturbed electron density. This is given by:
\begin{equation}\label{eq1}
\frac{\partial \Theta^{(2)}}{\partial \tau} + \bm{\hat{n}}.\bm{\nabla_x}\Theta^{(2)}
-\dot{\kappa}\Theta^{(2)} = 
-\dot{\kappa}\delta_e\left[\Theta_0^{(1)}-\Theta^{(1)}+\bm{\hat{n}}.\bm{V_b}
  -\frac{1}{2}\mathcal{P}_2(\bm{\hat{V_b}}.\bm{\hat{n}})\Pi^{(1)}\right],\nonumber
\end{equation}
where $\tau$ is conformal time and $\tau_0$ its value today. $\Theta = \Delta T/T = \Theta^{(1)}+\Theta^{(2)}+\hspace{4 pt}
\rm{higher}\hspace{4 pt} \rm{order}\hspace{4 pt} \rm{terms}$ is the fractional
perturbation of CMB temperature, superscripts indicate the order of
perturbation while subscripts denote the multipole moment. All other
perturbations are of first order and we will omit the superscript for
them. Vector quantities are in bold face and their magnitudes in normal
face with $\bm{\hat{}}$ denoting unit vectors. We have omitted the factor of $1/2$ usually multiplied with
the second order term \cite{bar1} for convenience. $\bm{\hat{n}}$ is the
unit vector along the line of sight, $\dot{\kappa}\equiv
d\kappa/d\tau=-\bar{n_e}\sigma_Ta$ is the mean differential optical depth
due to Compton scattering, $\sigma_T$ is the Thomson scattering cross
section and $a$ is the scale factor. We take the electron velocity to be equal
to the baryon velocity $\bm{V_b}$. $\mathcal{P}_2$ is the Legendre
polynomial of order 2 and 
$\Pi^{(1)}=\Theta_2^{(1)}+\Theta_{P0}^{(1)}+\Theta_{P2}^{(1)}$ is the
polarization term, subscript $P$ denoting the polarization field \cite{ma}.
We must caution
 that this partial equation is gauge dependent because $\delta_e$
depends on the gauge. We will be using conformal Newtonian gauge for
$\delta_e$. The combinations of terms multiplying $\delta_e$ is gauge invariant. All
perturbed quantities are  functions of $\tau$ and coordinates on spatial
slice $\bm{x}$. $\Theta$ is in addition a function of line of sight angle $\bm{\hat{n}}$.

Following standard  procedure \cite{cmbfast,bar1}, we
take the Fourier transform of Equation \ref{eq1} and integrate formally along
the line of sight.
\begin{eqnarray}
\Theta^{(2)}(\bm{k},\bm{\hat{n}},\tau_0) & = &
\int_0^{\tau_0}d\tau e^{ik\mu\left(\tau-\tau_0\right)}g(\tau)\int\frac{d^3\bm{k'}}{\left(2\pi\right)^3}\delta_e(\bm{k}-\bm{k'},\tau)\nonumber\\
&\times&\left[\Theta_0^{(1)}(\bm{k'},\tau)-\Theta^{(1)}(\bm{k'},\bm{\hat{n}},\tau)+
\bm{\hat{n}}.\bm{\hat{k'}}V_b(\bm{k'},\tau)
  -\frac{1}{2}\mathcal{P}_2(\bm{\hat{k'}}.\bm{\hat{n}})\Pi^{(1)}(\bm{k'},\tau)\right],\label{fourier}
\end{eqnarray}
where $g(\tau)=-\dot{\kappa}(\tau)e^{-\kappa(\tau)}$ is the visibility
function and $\kappa(\tau)\equiv\int_{\tau}^{\tau_0}d\tau ' \bar{n_e}(\tau
')\sigma_Ta(\tau ')$. Also $\bm{V_b}(\bm{k'},\tau)=\bm{\hat{k'}}V_b(\bm{k'},\tau)$.
We now take the spherical harmonic transform of Equation \ref{fourier} to
get the multipole moments, $\Theta^{(2)}_{\ell m}$.
\begin{eqnarray}
\Theta^{(2)}_{\ell m}(\bm{k},\tau_0) & = & \int
\Theta^{(2)}(\bm{k},\bm{\hat{n}},\tau_0) Y_{\ell
  m}^{\ast}(\bm{\hat{n}})d\bm{\hat{n}}\nonumber
\end{eqnarray}

This integral can be performed after decomposing $\Theta^{(1)}$ into
multipole moments,
$\Theta^{(1)}(\bm{k'},\bm{\hat{n}},\tau)=\sum_{\ell''}(-i)^{\ell ''}(2\ell
''+1)\mathcal{P}_{\ell ''}(\bm{\hat{n}}.\bm{\hat{k'}})\Theta^{(1)}_{\ell
  ''}(\bm{k'},\tau)$ and using relations between exponential, spherical
harmonics, spherical Bessel functions and Legendre polynomials \cite{var}.
Note that  $\Theta_0^{(1)}$, which is the dominant term in the multipole
expansion of $\Theta^{(1)}$,   will cancel out. We will see later that the
dipole term partially cancels the effect of ($V_b$), the Vishniac term. So only
$\ell \ge 2$ modes in $\Theta^{(1)}$, which are expected to be small
compared to monopole,  will contribute to the bispectrum.
The result is:
\begin{eqnarray}
&&\Theta^{(2)}_{\ell m}(\bm{k},\tau_0)  =  \int_0^{\tau_0}d\tau g(\tau)\int
\frac{d^3\bm{k'}}{\left(2\pi\right)^3}\delta_e(\bm{k}-\bm{k'},\tau)\biggl[ \nonumber\\
&& - (4\pi)^2\sum_{\ell 'm' \ell ''\neq 0 m''}i^{\ell '}(-i)^{\ell ''}\sqrt{\frac{(2\ell
    '+1)(2\ell ''+1)}{4\pi(2\ell +1)}}C^{\ell 0}_{\ell ' 0 \ell ''
  0}C^{\ell m}_{\ell ' m' \ell '' m''}j_{\ell '}[k(\tau-\tau_0)]Y_{\ell '
  m'}^{\ast}(\bm{\hat{k}})Y_{\ell '' m''}^{\ast}(\bm{\hat{k'}})\Theta_{\ell
  ''}^{(1)}(\bm{k'},\tau)\nonumber\\
&& +\frac{(4\pi)^2}{3}\sum_{\ell 'm' m''}i^{\ell '}\sqrt{\frac{(2\ell
    '+1)3}{4\pi(2\ell +1)}}C^{\ell 0}_{\ell ' 0 1
  0}C^{\ell m}_{\ell ' m' 1 m''}j_{\ell '}[k(\tau-\tau_0)]Y_{\ell '
  m'}^{\ast}(\bm{\hat{k}})Y_{1
  m''}^{\ast}(\bm{\hat{k'}})V_b(\bm{k'},\tau)\nonumber\\
\label{theta2lm}&& -\frac{1}{2}\frac{(4\pi)^2}{5}\sum_{\ell 'm' m''}i^{\ell '}\sqrt{\frac{(2\ell
    '+1)5}{4\pi(2\ell +1)}}C^{\ell 0}_{\ell ' 0 2
  0}C^{\ell m}_{\ell ' m' 2 m''}j_{\ell '}[k(\tau-\tau_0)]Y_{\ell '
  m'}^{\ast}(\bm{\hat{k}})Y_{2
  m''}^{\ast}(\bm{\hat{k'}})\Pi^{(1)}(\bm{k'},\tau)
\biggr]\\
&&\equiv \int_0^{\tau_0}d\tau g(\tau)\int
\frac{d^3\bm{k'}}{\left(2\pi\right)^3}\delta_e(\bm{k}-\bm{k'},\tau)S^{\ell m}(\bm{k},\bm{\hat{k'}},\bm{k'},\tau)\nonumber
\end{eqnarray}

$C^{\ell m}_{\ell ' m' \ell '' m''}$ are Clebsch-Gordon coefficients,
$j_{\ell}$ are spherical Bessel functions. The sums are over all allowed values
of $\ell m$ with the exceptions explicitly specified. The last line defines
the function $S^{\ell m}$. Its arguments are written so that we can keep track
of the part, $\bm{k'}$, that statistical variables like temperature
anisotropy depend on from the part that deterministic functions depend on,
$\bm{\hat{k'}}$, separately.
\end{section}

\begin{section}{Bispectrum}
We can now use Equation \ref{theta2lm} to calculate the bispectrum. This is
defined as:
\begin{eqnarray}\label{bispec}
B_{m_1m_2m_3}^{\ell_1\ell_2\ell_3}=\langle
a_{\ell_1m_1}^{(1)}(\bm{x},\tau_0)a_{\ell_2m_2}^{(1)}(\bm{x},\tau_0)a_{\ell_3m_3}^{(2)}(\bm{x},\tau_0)\rangle
+ \hspace{4 pt}2\hspace{4 pt}\rm{permutations},
\end{eqnarray}
where $a_{\ell m}(\bm{x},\tau_0)$ are the coefficients in the spherical
harmonic expansion of the corresponding temperature anisotropy. $\langle
\rangle$ denotes the ensemble average. At second
order they are just the Fourier transform of $\Theta^{(2)}_{\ell
  m}(\bm{k},\tau_0)$ while at first order they can be computed from
$\Theta^{(1)}_{\ell}(\bm{k},\tau_0)$.
\begin{eqnarray}
a_{\ell m}^{(2)}(\bm{x},\tau_0)&=&\int\frac{d^3\bm{k}}{(2\pi)^3}e^{i\bm{k}.\bm{x}}
\Theta^{(2)}_{\ell m}(\bm{k},\tau_0)\nonumber\\
a_{\ell m}^{(1)}(\bm{x},\tau_0)&=&4\pi\int\frac{d^3\bm{k}}{(2\pi)^3}e^{i\bm{k}.\bm{x}}
(-i)^{\ell}\Theta^{(1)}_{\ell}(\bm{k},\tau_0)Y_{\ell m}^{\ast}(\bm{\hat{k}})\nonumber
\end{eqnarray}
We can now calculate the first term of the bispectrum in Equation
\ref{bispec}.
\begin{eqnarray}
\langle 1,1,2  \rangle &\equiv &
\langle
a_{\ell_1m_1}^{(1)}(\bm{x},\tau_0)a_{\ell_2m_2}^{(1)}(\bm{x},\tau_0)a_{\ell_3m_3}^{(2)}(\bm{x},\tau_0)\rangle\nonumber\\
&=& (4\pi)^2\int
\frac{d^3\bm{k_1}}{(2\pi)^3}\frac{d^3\bm{k_2}}{(2\pi)^3}\frac{d^3\bm{k_3}}{(2\pi)^3}e^{i(\bm{k_1}+\bm{k_2}+\bm{k_3}).\bm{x}}(-i)^{\ell_1+\ell_2}Y_{\ell_1m_1}^{\ast}(\bm{\hat{k_1}})Y_{\ell_2m_2}^{\ast}(\bm{\hat{k_2}})\nonumber\\
&&\int_0^{\tau_0}d\tau g(\tau)\int
\frac{d^3\bm{k'}}{(2\pi)^3}\langle \delta_e(\bm{k_3}-\bm{k'},\tau)S^{\ell_3m_3}(\bm{k_3},\bm{\hat{k'}},\bm{k'},\tau)\Theta^{(1)}_{\ell_1}(\bm{k_1},\tau_0)\Theta^{(1)}_{\ell_2}(\bm{k_2},\tau_0)\rangle\nonumber\\
\end{eqnarray}
We can write each term in the ensemble average as a transfer function times
initial gravitational potential perturbation. Thus,
\begin{eqnarray}
\delta_e(\bm{k_3}-\bm{k'},\tau)&=&\Phi_i(\bm{k_3}-\bm{k'})\delta_e(|\bm{k_3}-\bm{k'}|,\tau)\nonumber\\
\Theta^{(1)}_{\ell_1}(\bm{k_1},\tau_0)&=&\Phi_i(\bm{k_1})\Theta^{(1)}_{\ell_1}({k_1},\tau_0)\nonumber\\
S^{\ell_3m_3}(\bm{k_3},\bm{\hat{k'}},\bm{k'},\tau)&=&\Phi_i(\bm{k'})S^{\ell_3m_3}(\bm{k_3},\bm{\hat{k'}},k',\tau)\nonumber\\
\langle \Phi_i(\bm{k_1})\Phi_i(\bm{k_2})\rangle &=& (2\pi)^3\delta^3(\bm{k_1}+\bm{k_2})P(k_1)\nonumber
\end{eqnarray}
We are using same symbols for statistical variables and their deterministic
transfer function counterparts, with arguments determining which one we
mean. $\delta^3$ is the three dimensional Dirac delta distribution and
$P(k)=2\pi^2/k^3$ is the initial power spectrum. Since we assume the initial
perturbation to be Gaussian, the 4-point ensemble average can be decomposed into
2-point ensemble averages.
\begin{eqnarray}
\langle
\Phi_i(\bm{k_3}-\bm{k'})\Phi_i(\bm{k'})\Phi_i(\bm{k_1})\Phi_i(\bm{k_2})\rangle
&=&\langle
\Phi_i(\bm{k_3}-\bm{k'})\Phi_i(\bm{k'}))\rangle\langle\Phi_i(\bm{k_1})\Phi_i(\bm{k_2})\rangle\nonumber\\
&+&\langle 
\Phi_i(\bm{k_3}-\bm{k'})\Phi_i(\bm{k_1}))\rangle\langle\Phi_i(\bm{k'})\Phi_i(\bm{k_2})\rangle\nonumber\\
&+&\langle 
\Phi_i(\bm{k_3}-\bm{k'})\Phi_i(\bm{k_2}))\rangle\langle\Phi_i(\bm{k'})\Phi_i(\bm{k_1})\rangle\nonumber\\
&=&(2\pi)^6\delta^3(\bm{k_3})P(k')\delta^3(\bm{k_1}+\bm{k_2})P(k_1)\nonumber\\
&+&(2\pi)^6\delta^3(\bm{k_3}+\bm{k_1}-\bm{k'})P(k_1)\delta^3(\bm{k_2}+\bm{k'})P(k_2)\nonumber\\
\label{wick}&+&(2\pi)^6\delta^3(\bm{k_3}+\bm{k_2}-\bm{k'})P(k_2)\delta^3(\bm{k_1}+\bm{k'})P(k_1)
\end{eqnarray}
First term in Equation \ref{wick} contributes only for $\bm{k_3}=0$, it is
a product of monopole and power spectrum and is unobservable. The other two
terms are identical with $\bm{k_1},\bm{k_2}$ terms interchanged. So we need
consider only one of these. Denoting the two terms by superscript $(1,2)$
and $(2,1)$  we can write the first term of
the bispectrum as:
\begin{eqnarray}
\langle 1,1,2  \rangle &=& \langle 1,1,2  \rangle^{(1,2)}+\langle 1,1,2
  \rangle^{(2,1)},\nonumber\\
\langle 1,1,2  \rangle^{(1,2)}& =&  (4\pi)^2\int
\frac{d^3\bm{k_1}}{(2\pi)^3}\frac{d^3\bm{k_2}}{(2\pi)^3}\frac{d^3\bm{k_3}}{(2\pi)^3}e^{i(\bm{k_1}+\bm{k_2}+\bm{k_3}).\bm{x}}(-i)^{\ell_1+\ell_2}Y_{\ell_1m_1}^{\ast}(\bm{\hat{k_1}})Y_{\ell_2m_2}^{\ast}(\bm{\hat{k_2}})\nonumber\\
&& \int_0^{\tau_0}d\tau g(\tau)\int
\frac{d^3\bm{k'}}{(2\pi)^3}\delta_e(k_1,\tau)S^{\ell_3m_3}(\bm{k_3},\bm{\hat{k'}},k_2,\tau)\Theta^{(1)}_{\ell_1}(k_1,\tau_0)\Theta^{(1)}_{\ell_2}(k_2,\tau_0)\nonumber\\
&&(2\pi)^6\delta^3(\bm{k_3}+\bm{k_1}-\bm{k'})P(k_1)\delta^3(\bm{k_2}+\bm{k'})P(k_2)\nonumber\\
&=& (4\pi)^2(2\pi)^3\int_0^{\tau_0}d\tau g(\tau)\int
\frac{d^3\bm{k_1}}{(2\pi)^3}\frac{d^3\bm{k_2}}{(2\pi)^3}\frac{d^3\bm{k_3}}{(2\pi)^3}(-i)^{\ell_1+\ell_2}Y_{l_1m_1}^{\ast}(\bm{\hat{k_1}})Y_{\ell_2m_2}^{\ast}(\bm{\hat{k_2}})\nonumber\\
&&P(k_1)P(k_2) 
\delta_e(k_1,\tau)S^{\ell_3m_3}(\bm{k_3},\bm{\hat{-k_2}},k_2,\tau)\Theta^{(1)}_{\ell_1}(k_1,\tau_0)\Theta^{(1)}_{\ell_2}(k_2,\tau_0)\delta^3(\bm{k_1}+\bm{k_2}+\bm{k_3})\nonumber\\\label{112}
\end{eqnarray}
In the last step we have used one of the Dirac delta distributions to
integrate over $\bm{k'}$.
To proceed further we use the representation of Dirac delta distribution as
Fourier transform of unity and the expansion of exponential function in
spherical harmonics.
\begin{eqnarray}
\delta^3(\bm{k_1}+\bm{k_2}+\bm{k_3})&=&\int
\frac{d^3{\bm{r}}}{(2\pi)^3}e^{i(\bm{k_1}+\bm{k_2}+\bm{k_3}).\bm{r}}\nonumber\\
e^{i(\bm{k}.\bm{r})}
&=&\label{dirac}4\pi\sum_{\ell,m}i^{\ell}j_{\ell}(kr)Y_{\ell m}^{\ast}(\bm{\hat{k}})Y_{\ell m}(\bm{\hat{r}})
\end{eqnarray}
Using Equations \ref{dirac} in \ref{112} we can perform all angular
integrals and all radial integrals except two which involve transfer
functions of perturbations and the line of sight integral. The integrals
involving spherical harmonics result in Wigner 3jm symbols which can then
be summed using, for example, formulas tabulated in \cite{var}.
The result after performing these integrals is:
\begin{eqnarray}
\label{b12}\langle 1,1,2  \rangle^{(1,2)}&=&\sqrt{\frac{(2\ell_1 +1)(2\ell_2 +1)
(2\ell_3 +1)}{4\pi}}\left(
\begin{array}{lcr}
\ell_1 & \ell_2 & \ell_3 \\
  0 & 0 & 0
\end{array}
\right)
\left(
\begin{array}{lcr}
\ell_1 & \ell_2 & \ell_3 \\
  m_1 & m_2 & m_3
\end{array}
\right)
\int_0^{\tau_0}d\tau
g(\tau)B_{\delta\Theta}^{\ell_1}(\tau)B_{\Theta\Theta}^{\ell_2}(\tau)\nonumber\\\hspace{4
  pt}\\
B_{\delta\Theta}^{\ell_1}(\tau)&=&\frac{2}{\pi}\int
k_1^2dk_1P(k_1)\Theta_{\ell_1}^{(1)}(k_1,\tau_0)\delta_e(k_1,\tau)j_{\ell_1}[k_1(\tau_0-\tau)]\nonumber\\
B_{\Theta\Theta}^{\ell_2}(\tau)&=&\frac{2}{\pi}\int
k_2^2dk_2P(k_2)\Theta_{\ell_2}^{(1)}(k_2,\tau_0)\biggl[\nonumber\\
&&-\sum_{\ell ''\geq 1, \ell_2'} i^{\ell ''+\ell_2 + \ell_2 '}(-1)^{\ell_2}(2\ell
  ''+1)(2\ell_2 '+1)\left(
\begin{array}{lcr}
\ell_2 ' & \ell_2 & \ell '' \\
  0 & 0 & 0
\end{array}
\right)^2 
\Theta_{\ell ''}^{(1)}(k_2,\tau)j_{\ell_2'}[k_2(\tau_0-\tau)]\nonumber\\
&& + i V_b(k_2,\tau)j'_{\ell_2}
  [k_2(\tau_0-\tau)]
 +\frac{1}{4}\Pi^{(1)}(k_2,\tau)\left\{3j''_{\ell_2}[k_2(\tau_0-\tau)]+j_{\ell_2}\left[k_2(\tau_0-\tau)\right]\right\}
\biggr]\nonumber\\
&=&\frac{2}{\pi}\int
k_2^2dk_2P(k_2)\Theta_{\ell_2}^{(1)}(k_2,\tau_0)\biggl[\nonumber\\
&&-\sum_{\ell ''\geq 2, \ell_2'} i^{\ell ''+\ell_2 + \ell_2 '}(-1)^{\ell_2}(2\ell
  ''+1)(2\ell_2 '+1)\left(
\begin{array}{lcr}
\ell_2 ' & \ell_2 & \ell '' \\
  0 & 0 & 0
\end{array}
\right)^2 
\Theta_{\ell ''}^{(1)}(k_2,\tau)j_{\ell_2'}[k_2(\tau_0-\tau)]\nonumber\\
& + &\left[\theta_b(k_2,\tau)-\theta_{\gamma}(k_2,\tau)\right]\frac{j'_{\ell_2}
  [k_2(\tau_0-\tau)]}{k_2}\nonumber\\
\label{btt}& +&\frac{1}{4}\Pi^{(1)}(k_2,\tau)\left\{3j''_{\ell_2}[k_2(\tau_0-\tau)]+j_{\ell_2}\left[k_2(\tau_0-\tau)\right]\right\}
\biggr]
\end{eqnarray}
In the last step we have defined $iV_b=\theta_b/k$ and
$\theta_{\gamma}=3k\Theta_1$ and evaluated the sum over $\ell_2'$ explicitly for
$\Theta_1$. It can be seen from this expression that the effect of Vishniac
term $\theta_b$ is partly cancelled out by $\theta_{\gamma}$.
In this form the gauge invariance of $B_{\Theta\Theta}^{\ell_2}$ is also
apparent. In arriving at these expressions we have also used the identity
$j_{\ell}(-x)=(-1)^{\ell}j_{\ell}(x)$. The prime on the Bessel functions
denotes the derivative with respect to the argument.

We can now write down the final expression for the angular averaged
bispectrum defined by:
\begin{eqnarray}
B^{\ell_1\ell_2\ell_3}&=&\sum_{m_1m_2m_3}\left(
\begin{array}{lcr}
\ell_1 & \ell_2 & \ell_3 \\
  m_1 & m_2 & m_3
\end{array}
\right)
B_{m_1m_2m_3}^{\ell_1\ell_2\ell_3}\nonumber\\
&=&\sqrt{\frac{(2\ell_1 +1)(2\ell_2 +1)
(2\ell_3 +1)}{4\pi}}\left(
\begin{array}{lcr}
\ell_1 & \ell_2 & \ell_3 \\
  0 & 0 & 0
\end{array}
\right)
\int_0^{\tau_0}d\tau
g(\tau)\biggl[ B_{\delta\Theta}^{\ell_1}(\tau)B_{\Theta\Theta}^{\ell_2}(\tau)\nonumber\\
&&+
B_{\delta\Theta}^{\ell_2}(\tau)B_{\Theta\Theta}^{\ell_1}(\tau)+
B_{\delta\Theta}^{\ell_2}(\tau)B_{\Theta\Theta}^{\ell_3}(\tau)+
B_{\delta\Theta}^{\ell_3}(\tau)B_{\Theta\Theta}^{\ell_2}(\tau)+
B_{\delta\Theta}^{\ell_1}(\tau)B_{\Theta\Theta}^{\ell_3}(\tau)+
B_{\delta\Theta}^{\ell_3}(\tau)B_{\Theta\Theta}^{\ell_1}(\tau)
\biggr]\nonumber\\
\end{eqnarray}

\end{section}
\begin{section}{Primordial Non-Gaussianity of local type}
We will compare our results with the bispectrum from primordial
non-Gaussianity of local type. This is given by \cite{komatsu01,review}:
\begin{eqnarray}
B^{\ell_1\ell_2\ell_3}_{prim}&=&2
\sqrt{\frac{(2\ell_1 +1)(2\ell_2 +1)
(2\ell_3 +1)}{4\pi}}\left(
\begin{array}{lcr}
\ell_1 & \ell_2 & \ell_3 \\
  0 & 0 & 0
\end{array}
\right)
\int_0^{\tau_0}d\tau
(\tau_0-\tau)^2\biggl[ \beta_{\ell_1}(\tau)\beta_{\ell_2}(\tau)\alpha_{\ell_3}(\tau)\nonumber\\
&&+\beta_{\ell_2}(\tau)\beta_{\ell_3}(\tau)\alpha_{\ell_1}(\tau)
+\beta_{\ell_3}(\tau)\beta_{\ell_1}(\tau)\alpha_{\ell_2}(\tau)\biggr],\nonumber\\
\beta_{\ell}(\tau)&=&\frac{2}{\pi}\int k^2dk P(k)\Theta_{\ell}(k,\tau_0)j_{\ell}[k(\tau_0-\tau)],\nonumber\\
\alpha_{\ell}(\tau)&=&\frac{2}{\pi}\int k^2dk f_{NL}\Theta_{\ell}(k,\tau_0)j_{\ell}[k(\tau_0-\tau)],
\end{eqnarray}
where $f_{NL}$ is the non-Gaussianity parameter defined by the following
form for the primordial potential,
$\Phi_i(\bm{x})=\Phi_L(\bm{x})+f_{NL}\left(\Phi^2_L(\bm{x})-\left\langle\Phi^2_L(\bm{x})\right\rangle\right)$
with $\Phi_L(\bm{x})$ Gaussian. Note that the expression for $\beta_{\ell}$
is similar to $B_{\delta\Theta}^{\ell}$ and $B_{\Theta\Theta}^{\ell}$,
the difference being the additional modulation by the terms at recombination
in the later case. As we will see later,  $\alpha_{\ell}$ is similar in
shape to the visibility function $g({\tau})$ but peaks at an earlier time.
All plots and results are for $f_{NL}=1$.
\end{section}
\begin{section}{Numerical calculation and Results}
We calculate $\delta_e$ in conformal Newtonian gauge using DRECFAST \cite{novos}. All
other first order terms are calculated using CMBFAST \cite{cmbfast}. In
particular $\Theta_{\ell ''}(k,\tau)$ is given by the line of sight
integral:
\begin{equation}
\Theta_{\ell ''}(k,\tau) = e^{\kappa(\tau)}\int_0^{\tau}d\tau
'S^{(1)}(k,\tau ')j_{\ell ''}[k(\tau - \tau ')]
\end{equation}
Here $S^{(1)}(k,\tau ')$ is the usual first order source term. Since we are
evaluating the transfer function at $\tau < \tau_0$, we get an extra factor
of $e^{\kappa(\tau)}$, otherwise this is same as the standard line of sight
formula \cite{cmbfast}.
 $\Theta_{\ell ''}$ becomes smaller with increasing $\ell ''$
 and we cut off the sum in Equation \ref{btt} at $\ell '' = 30$. This is accurate for $\tau <
 1000 \rm{Mpc}$ which is sufficient for the present calculation since  the
 visibility $g(\tau)$ is non-negligible only for  $240\rm{Mpc} \lesssim
 \tau \lesssim 800 \rm{Mpc}$ (Figure \ref{alphaeps}). Wigner 3jm symbols
 are calculated using the code by Gordon and Schulten \cite{gordon} which
 is publicly available at SLATEC common mathematical library \cite{slatec}.

\begin{figure}
\includegraphics{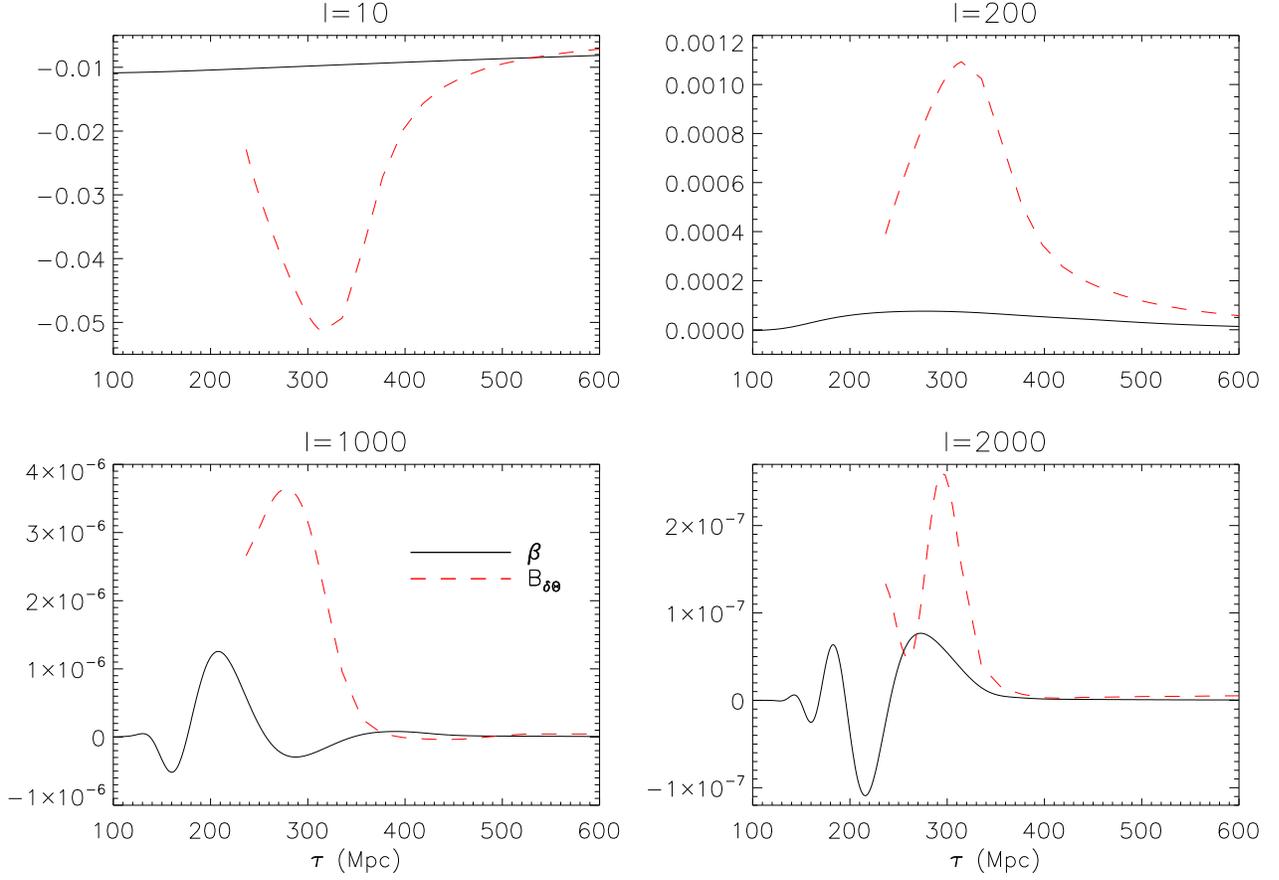}
\caption{\label{betaeps}$\beta_{\ell}(\tau)$ and
  $B_{\delta\Theta}^{\ell}(\tau)$ is shown as a function of $\tau$ for
  several values of $\ell$. } 
\end{figure}

\begin{figure}
\includegraphics{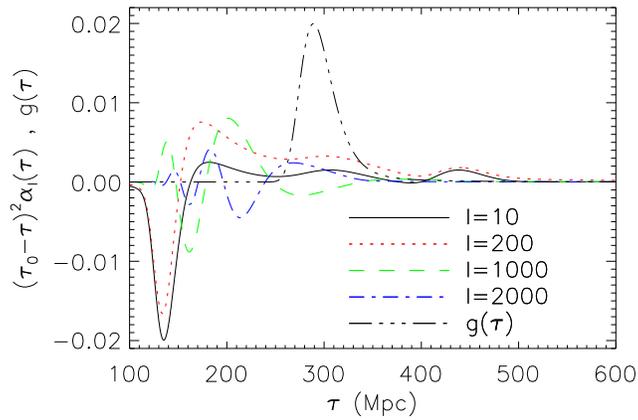}
\caption{\label{alphaeps}$(\tau_0-\tau)^2\alpha_{\ell}(\tau)$ for
  several values of $\ell$ and
  the visibility function $g(\tau)$  as a function of conformal time
  $\tau$. $(\tau_0-\tau)^2\alpha_{\ell}(\tau)$ peaks  earlier than
 $g(\tau)$.}
\end{figure}

\begin{figure}
\includegraphics{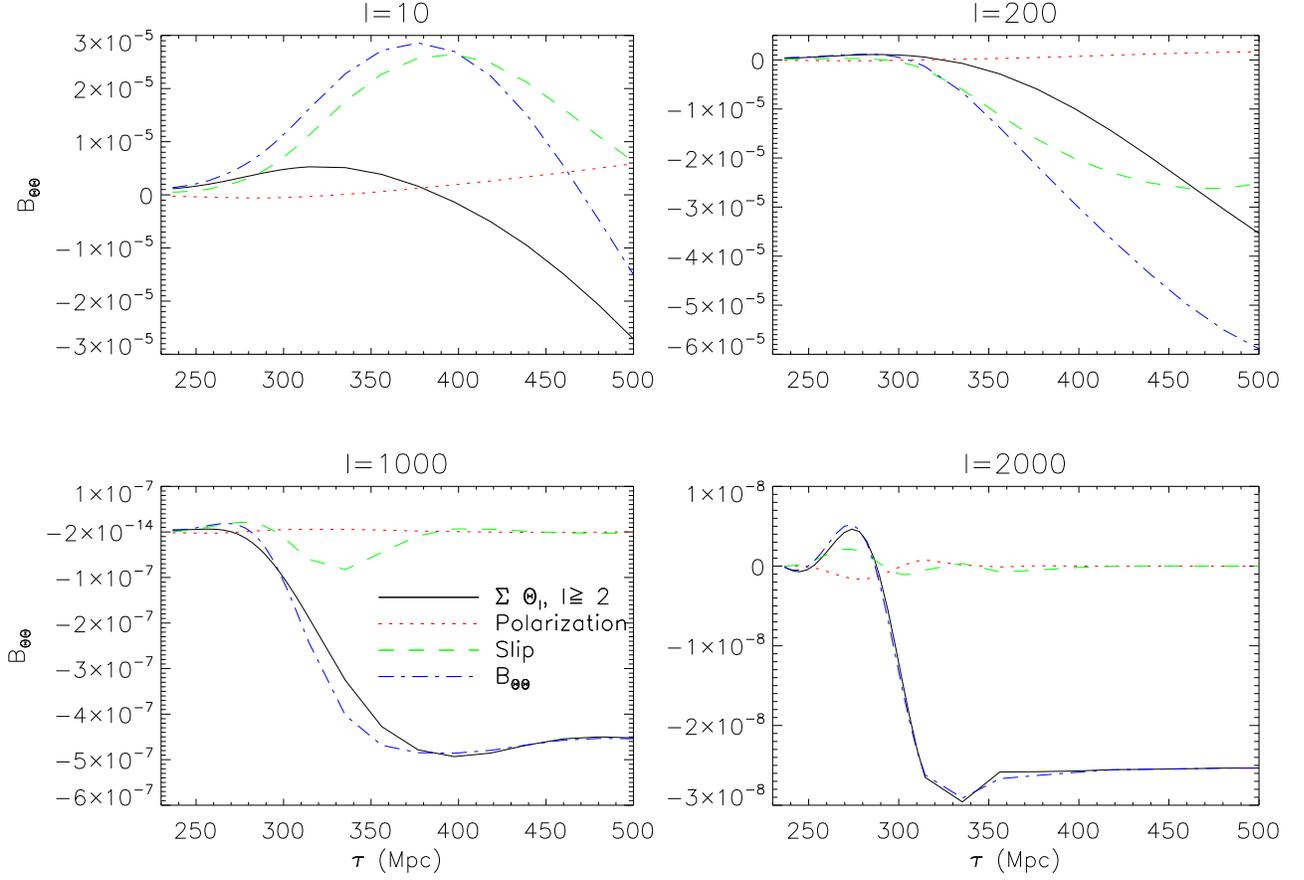}
\caption{\label{btteps} $B_{\Theta\Theta}^{\ell}(\tau)$ is shown for several
  values of $\ell$. Also shown are contributions from the polarization term
  $\Pi$, slip term $\theta_b-\theta_g$ and from all the other terms
  $\sum_{\ell \geq 2}\Theta^{(1)}_{\ell}$.
}
\end{figure}

\begin{figure}
\includegraphics{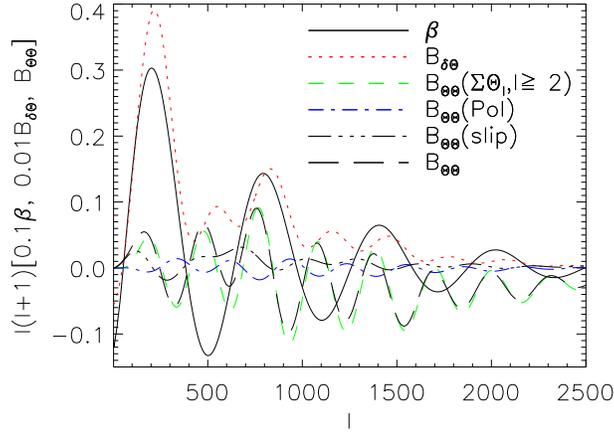}
\caption{\label{bleps} $0.1\times
  \ell(\ell+1)\beta_{\ell}(\tau_{\ast})$,$0.01\times
  \ell(\ell+1)B^{\ell}_{\delta\Theta}(\tau_{\ast})$,
  $\ell(\ell+1)B^{\ell}_{\Theta\Theta}(\tau_{\ast})$ and
  contributions to it from polarization, slip and rest of the terms is
  shown as a function of multipole moments $\ell$. Some of the functions have been scaled as specified above. }
\end{figure}
Figure \ref{betaeps} shows a comparison of $\beta_{\ell}(\tau)$ and
  $B_{\delta\Theta}^{\ell}(\tau)$. The modulation by $\delta_e$ results in
  shifting the peak to later times. Also visibility $g(\tau)$ 
  can be compared to primordial term $\alpha_{\ell}$. They are similar in
  magnitude but have a different shape (Figure \ref{alphaeps}).
  $B_{\Theta\Theta}^{\ell}$ is however much smaller in magnitude than the
  other terms, $B_{\delta\Theta}$ and $\beta_{\ell}$, as can be seen from
  Figures \ref{btteps} and \ref{bleps} at low $\ell$ but become comparable
  at high $\ell$. This results in a much smaller bispectrum
  from recombination at low $\ell$ compared to the primordial one.  Figures
  \ref{b10}, \ref{b200}, \ref{b1000} and \ref{b2000} show the absolute
  value of the bispectrum from the primordial
  non-Gaussianity with $f_{NL}=1$ and that due to inhomogeneous
  recombination for $\ell_3 = 10,200,1000,2000$ as a function of
  $\ell_1,\ell_2$. Z-axis is on linear scale while the color map is on log
  scale. They are almost identical at the peaks but differ considerably
  away from the peaks which occur when either $\ell_1$ or $\ell_2$ is
  equal to $\ell_3$ and the other is small, a signature of  the local
  nature of the non-Gaussianity. At low $\ell$ the amplitude $B^{\ell_1
    \ell_2 \ell_3}$  is much smaller than $B^{\ell_1
    \ell_2 \ell_3}_{prim}$ but they become comparable at high $\ell$. Their
  signs are however different and this will become apparent when we estimate
  the confusion in $f_{NL}$ due to $B^{\ell_1
    \ell_2 \ell_3}$.
\begin{figure}
\includegraphics{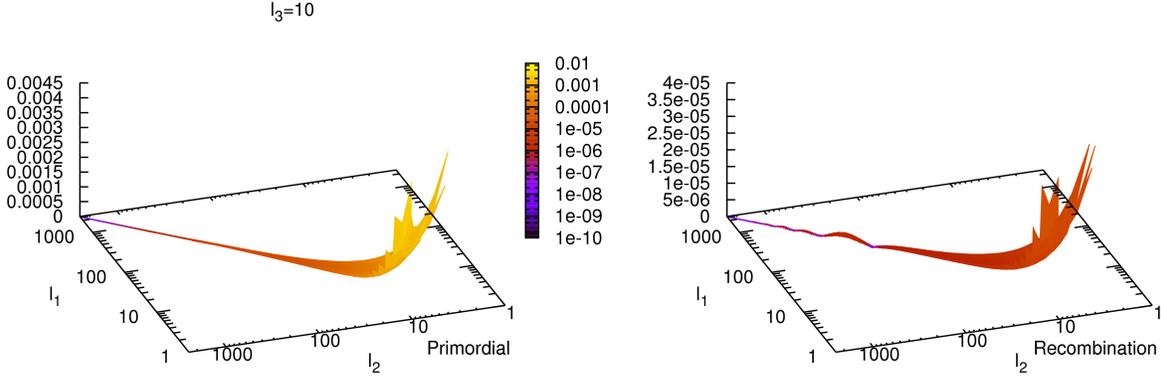}
\caption{\label{b10} Absolute value of $B^{\ell_1
    \ell_2 \ell_3}_{prim}$ labeled ``Primordial'' and $B^{\ell_1
    \ell_2 \ell_3}$ labeled ``Recombination'' for $\ell_3=10$. Z axis
  is on linear scale while color plot shows the same on log scale.
}
\end{figure}

\begin{figure}
\includegraphics{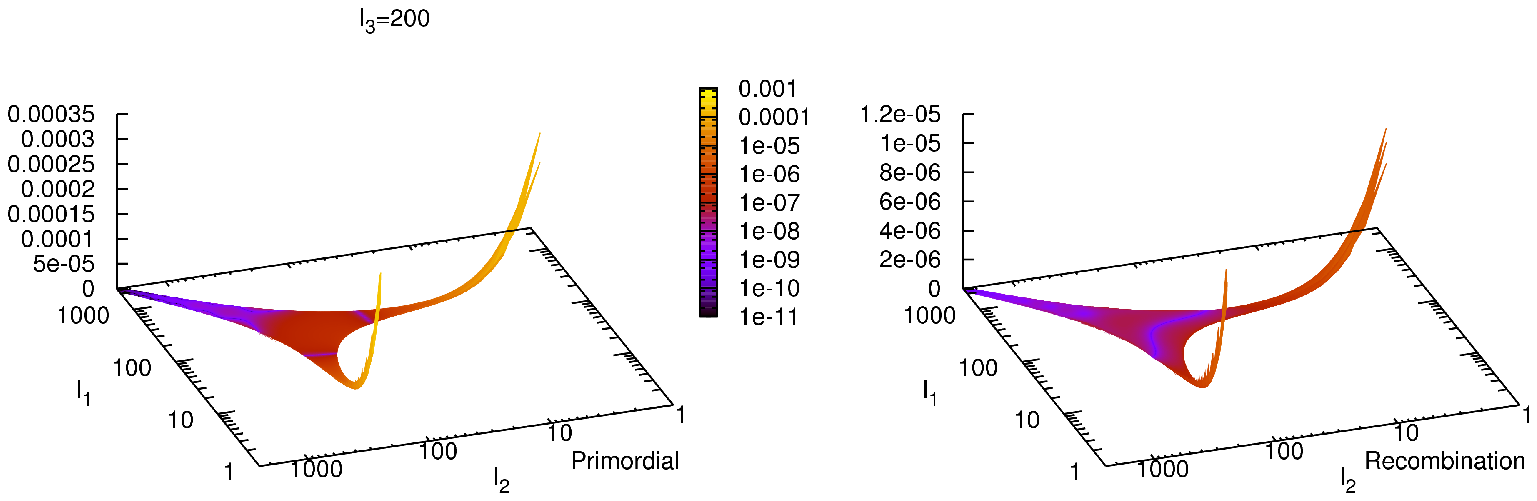}
\caption{\label{b200} Absolute value of $B^{\ell_1
    \ell_2 \ell_3}_{prim}$ labeled ``Primordial'' and $B^{\ell_1
    \ell_2 \ell_3}$ labeled ``Recombination''  for $\ell_3=200$. Z axis
  is on linear scale while color plot shows the same on log scale.
}
\end{figure}

\begin{figure}
\includegraphics{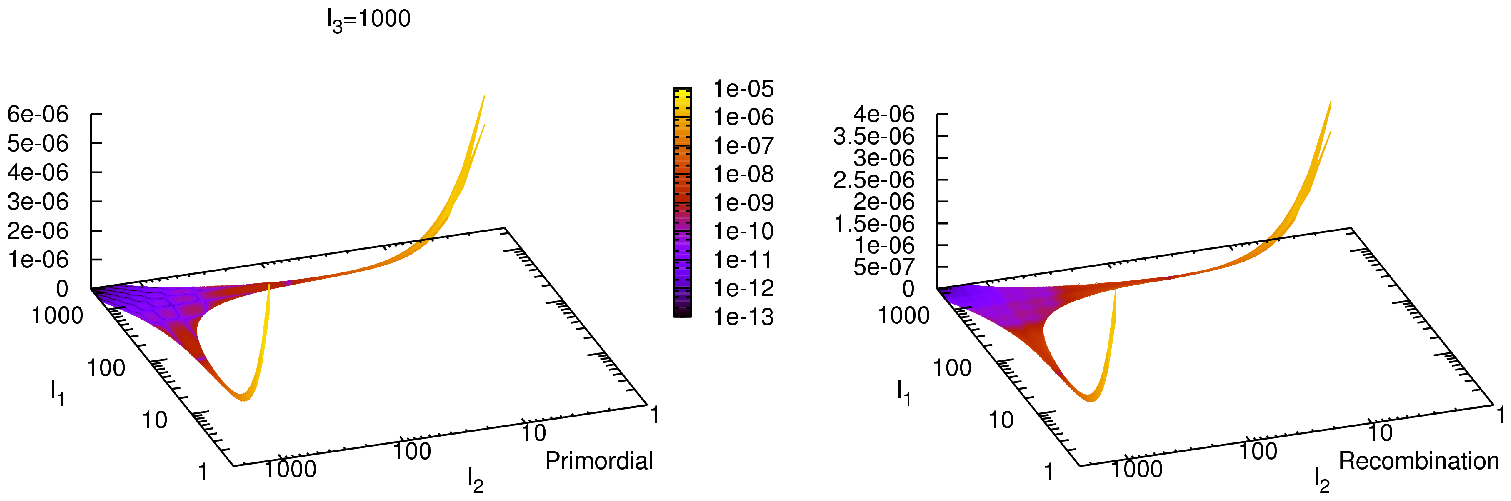}
\caption{\label{b1000} Absolute value of $B^{\ell_1
    \ell_2 \ell_3}_{prim}$ labeled ``Primordial'' and $B^{\ell_1
    \ell_2 \ell_3}$ labeled ``Recombination''  for $\ell_3=1000$. Z axis
  is on linear scale while color plot shows the same on log scale.
}
\end{figure}

\begin{figure}
\includegraphics{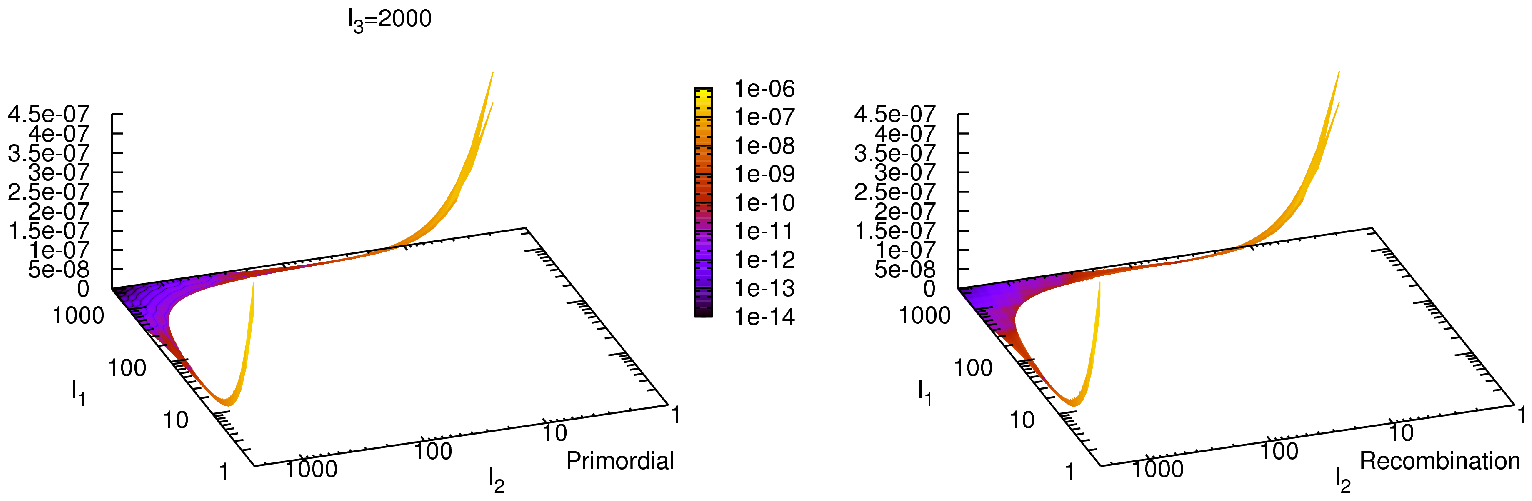}
\caption{\label{b2000} Absolute value of $B^{\ell_1
    \ell_2 \ell_3}_{prim}$ labeled ``Primordial'' and $B^{\ell_1
    \ell_2 \ell_3}$ labeled ``Recombination''  for $\ell_3=2000$. Z axis
  is on linear scale while color plot shows the same on log scale.
}
\end{figure}

To estimate the confusion to the estimate of $f_{NL}$ we follow \cite{fast}
and define the statistic
\begin{eqnarray}
S_{rec}&=&\sum_{\ell_1\leq \ell_2 \leq \ell_3}\frac{B^{\ell_1 \ell_2
    \ell_3}B^{\ell_1 \ell_2
      \ell_3}_{prim}}{C_{\ell_1}C_{\ell_2}C_{\ell_3}}\nonumber\\\label{fnl}
  &\simeq&f_{NL}\sum_{\ell_1\leq \ell_2 \leq \ell_3}\frac{(B^{\ell_1 \ell_2
    \ell_3}_{prim})^2}{C_{\ell_1}C_{\ell_2}C_{\ell_3}}
\end{eqnarray}
The result of solving Equation \ref{fnl} for $f_{NL}$ is shown in Figure
\ref{fnlfig} as a function of $\ell_{max}$, where $\ell_{max}$ is the
maximum value of $\ell$ included in the sum in Equation \ref{fnl}. As
expected from the examination of bispectra, $f_{NL}$ is small
and positive at low $\ell_{max}$ but $\sim -1$ at high $\ell_{max}$.

\begin{figure}
\includegraphics{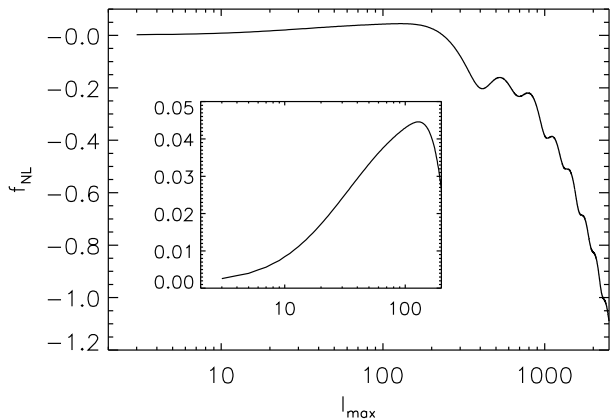}
\caption{\label{fnlfig} Comparison of primordial bispectrum from local type non-Gaussianity
with bispectrum  due to inhomogeneous recombination in terms of parameter
$f_{NL}$ as a function of $\ell_{max}$, the maximum $\ell$ mode
considered.}
\end{figure}
\end{section}

\begin{section}{Conclusions}
We have calculated the CMB bispectrum due to inhomogeneous
recombination. This was expected to be small because the combination of
terms multiplying $\delta_e$ is small. However calculations by Novosyadlyj
\cite{novos} showed that $\delta_e$ could be large and this suggested that the
CMB bispectrum could be non-negligible. Although it turns out to be small it
is still larger than what one might have expected from making an
estimate based on tight coupling or instantaneous recombination approximation \cite{bar2} and ignoring the
perturbations due to inhomogeneous recombination. This is especially evident
at high $\ell_{max}$. 
Also the bispectrum from recombination looks
remarkably like the local type primordial bispectrum, which is not entirely
unexpected since both arise due to product of two first order terms. 
Since the other second order terms in the Boltzmann equation \cite{bar1}  are expected to 
be larger than the ones we considered, our calculation motivates a full 
second order numerical calculation of these terms in order to assess their 
effect on future experiments such as Planck \cite{planck} and the level to which they 
cause confusion when probing for primordial non-Gaussianity.
\end{section}
\begin{acknowledgments}
 We thank Amit Yadav for help in the calculation of
the primordial bispectrum.
\end{acknowledgments}

\bibliographystyle{apsrev}
\bibliography{ng}
\end{document}